\pdfoutput=1
\documentclass{article}

\PassOptionsToPackage{numbers, compress}{natbib}
\usepackage[final]{neurips_2024}

\usepackage[utf8]{inputenc} % allow utf-8 input
\usepackage[T1]{fontenc}    % use 8-bit T1 fonts
\usepackage{hyperref}       % hyperlinks
\usepackage{url}            % simple URL typesetting
\usepackage{booktabs}       % professional-quality tables
\usepackage{amsfonts}       % blackboard math symbols
\usepackage{nicefrac}       % compact symbols for 1/2, etc.
\usepackage{microtype}      % microtypography
\usepackage{xcolor}         % colors

\usepackage{ragged2e}
\usepackage{array}
\newcolumntype{M}[1]{>{\RaggedRight\arraybackslash}m{#1}}

\usepackage{float}
\usepackage{enumitem}

\title{Gaps Between Research and Practice When
Measuring Representational Harms Caused by LLM-Based Systems}

\author{%
Emma Harvey$^{1}$\thanks{Work conducted during an internship at Microsoft Research} \quad Emily Sheng$^{2}$ \quad Su Lin Blodgett$^2$ \quad Alexandra Chouldechova$^2$ \\ \textbf{Jean Garcia-Gathright$^2$} \quad \textbf{Alexandra Olteanu}$^2$ \quad \textbf{Hanna Wallach}$^2$\\
$^1$Cornell University \quad $^2$Microsoft Research\\
\texttt{evh29@cornell.edu}
}

\begin{document}

\maketitle

\setcounter{footnote}{0}

\begin{abstract}
To facilitate the measurement of representational harms caused by large language model (LLM)-based systems, the NLP research community has produced and made publicly available numerous measurement instruments, including tools, datasets, metrics, benchmarks, annotation instructions, and other techniques. However, the research community lacks clarity about whether and to what extent these instruments meet the needs of practitioners tasked with developing and deploying LLM-based systems in the real world, and how these instruments could be improved. Via a series of semi-structured interviews with practitioners in a variety of roles in different organizations, we identify four types of challenges that prevent practitioners from effectively using publicly available instruments for measuring representational harms caused by LLM-based systems: (1) challenges related to using publicly available measurement instruments; (2) challenges related to doing measurement in practice; (3) challenges arising from measurement tasks involving LLM-based systems; and (4) challenges specific to measuring representational harms. Our goal is to advance the development of instruments for measuring representational harms that are well-suited to practitioner needs, thus better facilitating the responsible development and deployment of LLM-based systems.\looseness=-1
\end{abstract}

\section{Introduction}
As large language model (LLM)-based systems become increasingly widespread, so too has their potential to cause \textit{representational harms} \citep{barocas2017, crawford2017trouble}, which occur when a system ``represents some social groups in a less favorable light than others, demeans them, or fails to recognize their existence altogether'' \cite{blodgett_language_2020}. Representational harms are abstract concepts that cannot be measured directly \cite{jacobs_measurement_2021}---yet measuring such harms is important, as they can cause tangible negative outcomes, e.g., through the entrenchment of harmful social hierarchies, which may affect people's belief systems and psychological states \cite{corvi_representational_2024, wang2023measuring, chien_beyond_2024}. To facilitate the measurement of representational harms, the NLP research community has produced and made publicly available numerous \textit{measurement instruments},\footnote{By \textit{publicly available}, we mean that a measurement instrument has been made available on the internet or via an academic publication for others to use or adapt, potentially subject to licensing considerations.} including tools \citep[e.g.,][]{lees_new_2022, bucinca_aha_2023}, datasets \citep[e.g.,][]{zhao2024wildchat, gehman_realtoxicityprompts_2020, hada_fifty_2023, hartvigsen_toxigen_2022, sandoval_rose_2023}, metrics \citep[e.g.,][]{bommasani_trustworthy_2022, caliskan_semantics_2017, bolukbasi_man_2016, sheng_woman_2019, maheshwari_fair_2023}, benchmarks (consisting of both datasets and metrics) \citep[e.g.,][]{dhamala_bold_2021, parrish_bbq_2022, nadeem_stereoset_2021, nangia_crows-pairs_2020, rudinger_gender_2018, zhao_gender_2018, dev_building_2023, dinan_multi-dimensional_2020, esiobu_robbie_2023, fleisig_fairprism_2023, gonen_automatically_2020}, annotation instructions \citep[e.g.,][]{magooda_framework_2023}, and other techniques \citep[e.g.,][]{hofmann_dialect_2024, ribeiro_beyond_2020, wang_measuring_2022}. However, the research community lacks clarity about whether and to what extent these instruments meet the needs of practitioners tasked with developing and deploying LLM-based systems in the real world, and how the instruments could be improved.\looseness=-1

Via a series of semi-structured interviews with practitioners (N\,=\,12) in a variety of roles in different organizations, we identify four types of challenges that prevent practitioners from effectively using publicly available instruments for measuring representational harms caused by LLM-based systems: (1) challenges related to \textit{using publicly available measurement instruments}; (2) challenges related to doing measurement \textit{in practice}; (3) challenges arising from measurement tasks \textit{involving LLM-based systems}; and (4) challenges specific to measuring \textit{representational harms}. Our goal is to advance the development of instruments for measuring representational harms that are well-suited to practitioner needs, thus better facilitating the responsible development and deployment of LLM-based systems.\looseness=-1

\section{Methods}
We conducted 12 semi-structured interviews with practitioners who reported that their work involved measuring representational harms caused by LLM-based systems. We recruited participants through our professional networks, social media, cold emails to individuals identified through LinkedIn and conference proceedings, and snowball sampling \cite{morgan2008snowball}. All interviews were one-hour long and were conducted virtually between June and August 2024. Participants provided informed consent before their interviews and received \$75 gift cards afterwards. The study was approved by our institution's IRB.\looseness=-1

\textit{Participants.} Participants held research (7/12), applied science (2/12), engineering (2/12), and consulting (1/12) roles at large tech companies (6/12), AI-focused startups (3/12), large non-tech companies (2/12), and AI-focused nonprofits (1/12). They described working on a variety of LLM-based systems, including search engines and chatbots, as well as on content moderation tools for LLMs.\looseness=-1

\textit{Interviews and analysis.} We asked participants to describe their roles and the LLM-based systems they worked on. We also asked them to walk us through an example of how they measured representational harms, noting the publicly available measurement instruments they used or considered using. We then asked them to reflect on their experiences with those instruments and to discuss any challenges. Although our sample size is relatively small (a common problem when conducting research on AI practitioners \cite{scheuerman_walled_2024}), we conducted interviews until we reached saturation, i.e., until multiple consecutive interviews did not uncover any new perspectives \cite{small_2009_how, hennink_sample_2022}. Finally, we analyzed the resulting interview transcripts using a thematic analysis with an inductive--deductive coding approach \citep{braun_using_2006, braun_reflecting_2019}.\looseness=-1 

\section{Results}
We identified four types of challenges that prevent practitioners from effectively using publicly available instruments for measuring representational harms caused by LLM-based systems.\looseness=-1

\textit{Challenges related to using publicly available measurement instruments.} Although prior work has identified a range of challenges related to using publicly available measurement instruments \cite[e.g.,][]{blodgett_stereotyping_2021, solaiman2024evaluatingsocialimpactgenerative}, participants primarily reported challenges related to \textit{validity} and \textit{specificity}. Almost all participants (11/12) mentioned issues of validity---i.e., whether a measurement instrument meaningfully measures what stakeholders think it measures \cite{jacobs_measurement_2021}---related to \textit{correctness} (e.g., tools produce inaccurate outputs, datasets contain mislabeled instances) and \textit{contestedness} (e.g., different instruments use different definitions of representational harms). Similarly, almost all participants (11/12) mentioned issues of specificity---i.e., whether a measurement instrument is sufficiently specific to a system, its use case(s), and its deployment context(s). Examples included datasets that are too generic to align with relevant use cases like customer service chats, and labels that are not sufficiently detailed (e.g., categories like ``hate speech'' are labeled, but details like the targeted social group are missing).\looseness=-1 

\textit{Challenges related to doing measurement in practice.} 
Particularly salient to our participants (and in line with considerable prior work
\citep[e.g.,][]{balayn__2023, berman_scoping_2024, beutel_putting_2019, deng_exploring_2022, deng_understanding_2023, deng_investigating_2023, gehrmann_repairing_2023, holstein_improving_2019, lee_landscape_2021, madaio_co-designing_2020, ojewale_towards_2024, quinonero_candela_disentangling_2023, raji_closing_2020, rakova_where_2021, richardson_towards_2021}) were challenges related to doing measurement \textit{in practice}---i.e., when working within the constraints surrounding products and services that are deployed to real users. Some participants (4/12) felt that their measurements needed to produce \textit{specific quality assurances}, and suggested that those assurances might be better achieved through software testing practices.
They also pointed to data licensing and security issues (3/12), the need to align with company-specific policies (2/12), a lack of time to find publicly available measurement instruments in the first place (2/12), and competitive pressures\footnote{Specifically, pressures to develop new measurement instruments in order to claim proprietary capabilities.} (1/12) as factors incentivizing them to build new measurement instruments rather than adopting existing ones.\looseness=-1

\textit{Challenges arising from measurement tasks involving LLM-based systems.} Because the datasets that are used to train LLMs are often unknown, it can be difficult to tell whether an LLM-based system that performs well on a benchmark has simply been trained using the benchmark data. Multiple participants (6/12) therefore expressed discomfort with using publicly available benchmarks and datasets, even if they are valid and specific to their needs. Some reported finding suitable publicly available benchmarks, but then using them only as inspiration to create new, internal benchmarks from scratch.\looseness=-1

\textit{Challenges specific to measuring representational harms.} Finally, participants distinguished representational harms from other types of harms. Specifically, many (9/12) felt that compared to other types of harms (e.g., privacy violations), representational harms required more and different information to measure. Participants felt that measuring representational harms required context (6/12), alignment on essentially constructed constructs (2/12), and social science expertise (2/12) that measuring other types of harms did not. Some participants (2/12) also felt that they faced less commercial incentive to measure representational harms compared to other types of harms (e.g., quality of service harms), causing them to limit their measurement efforts. For example, some participants reported that relevant stakeholders felt that aligned models were so unlikely to generate outright demeaning content that it was not worth measuring in the first place. Others found that if they did not know how to \textit{mitigate} certain representational harms, relevant stakeholders would not value the measurement.\looseness=-1

\section{Discussion}
The four types of challenges described above shed light on whether and to what extent publicly available instruments for measuring representational harms meet the needs of practitioners tasked with developing and deploying LLM-based systems in the real world. Future work should further investigate these types of challenges and take steps to address them by, for example, drawing on measurement theory from the social sciences \citep[e.g.,][]{adcock_measurement_2001, jacobs_measurement_2021} and pragmatic measurement~\citep{glasgow_pragmatic_2013,hand2016measurement} to \textit{improve instruments} for measuring representational harms, and on other fields like implementation science \citep[e.g.,][]{bauer_introduction_2015, bauer_implementation_2020, enkin_using_1998, rogers_diffusion_1962} to \textit{improve the uptake} of publicly available measurement instruments among practitioners. \looseness=-1

\section{Broader Impacts}
By making explicit the key challenges
that prevent practitioners from effectively using publicly available instruments for measuring representational harms caused by LLM-based systems, our work is intended to serve as a starting point to \textit{bridge gaps between research and practice}. We hope that our findings will serve as a foundation for future work on measurement instruments that are better suited to practitioners' needs, as well as work on the adoption of publicly available measurement instruments. \looseness=-1

\section{Limitations}
The primary limitation of our study is our small sample size. As is the case with many studies targeting technology workers \cite{scheuerman_walled_2024}, it was challenging to identify and recruit potential participants. Practitioners who held relevant roles often declined to speak with us due to NDAs or other confidentiality concerns. As a result, we were only able to interview $12$ practitioners, some of whom declined to answer certain questions in order to remain in compliance with their employers' NDAs.\looseness=-1

\appendix
\section{Appendix}
Our semi-structured interview guide is shown below. As is typical of a semi-structured interview process, not every participant was asked exactly the same questions in exactly the same order, and some participants were asked additional follow-up or clarifying questions based on the answers they provided. The interview questions were supplemented with a set of slides containing definitions of key terms that we screenshared with participants. The definitions are included in the script below. 

\subsection{Introductions [5 min]}
Welcome! Thank you so much for taking the time for this interview. Before we get started, I just want to quickly introduce myself, talk about the goals of this study, and give you a chance to ask any questions you might have. This research study is intended to understand gaps between research and practice in evaluating large language model (LLM)-based systems, with a focus on measuring harms, adverse impacts, or other undesirable behaviors. In this interview, I'll ask you to share your experiences with and opinions on such evaluations, without discussing confidential information. I will also record this interview for the purpose of creating a deidentified transcript. If you prefer that your video not be recorded, please feel free to turn your camera off at this time. In addition, if at any point you would like to skip a question, take a break, or end the interview, please feel free to do so.

Do you have any questions before we get started?

\subsection{Background [5 min]}
    \begin{itemize}[left=19pt]
        \item[\textbf{[IQ1]}] To start, please briefly describe your role, focusing on your professional experience as it relates to LLM-based systems. 
        \item[\textbf{[IQ2]}] Can you briefly describe the LLM-based system(s) that you have previously evaluated, currently evaluate, or plan to evaluate? 
    \end{itemize}
\subsection{Experience with measurement instruments for representational harms [15 min]}
    \begin{itemize}[left=19pt]
        \item[\textbf{[IQ3]}] Throughout this interview, I will be focusing primarily on representational harms, which occur when ``a system represents some social groups in a less favorable light than it represents other groups by stereotyping them, demeaning them, or failing to recognize their existence altogether.'' 
        \item[] What examples of representational harms caused by LLM-based systems are you aware of? 
        \item[] \textit{If interviewee was not familiar with representational harms, we provided the following examples:}
        \begin{itemize}[left=0pt]
        \item[--] LLMs might reinforce stereotypes, for example, by using the word ``nurse'' to refer to a female healthcare provider and the word ``doctor'' to refer to a male healthcare provider in otherwise identical contexts.
        \item[--] LLMs might generate slurs or derogatory language about a social group.
        \item[--] LLMs might erase a social group, for example, by only listing male athletes when a user asks for examples of talented soccer players, thus failing to recognize the existence of non-male soccer players.
        \end{itemize}
        \item[\textbf{[IQ4]}] Do your previous, current, or planned evaluation(s) of LLM-based system(s) involve measuring representational harms? 
    \item[\textbf{[IQ5]}] What types of representational harms are you measuring? 
    \item[\textbf{[IQ6]}] Can you walk me through, from start to finish, an example of how you measure representational harms? I'm especially interested in hearing about how you decided on your approach, whether you relied on existing, publicly available tools, benchmarks, datasets, metrics, annotation guidelines, and so on, or whether you decided to develop your own. 
    \item[] \textit{To allow for open-ended discussion, we did not provide participants with a specific definition of `measurement instruments'; rather, we provided the following examples of instruments:}
    \begin{itemize}[left=0pt]
    \item[--] An example of a tool is Perspective API.
    \item[--] An example of a benchmark is StereoSet, which includes a dataset of prompts that could elicit stereotyping content with corresponding metrics that measure the extent to which a language model produces stereotypes.
    \item[--] An example of a dataset is WildChat, which is a corpus of 1 million real user-ChatGPT interactions.
    \item[--] Examples of metrics are the Word and Sentence Embedding Association Tests (WEAT and SEAT), which measure whether “attribute words” (e.g. male, female) are disproportionately associated with a set of “target words” (e.g. different professions).
    \item[--] Annotation instructions are sets of instructions and examples for humans to use when annotating system outputs for particular properties.
    \item[--] An example of another type of instrument is Matched Guide Probing, a method adapted from sociolinguistics.
    \end{itemize}
\end{itemize}
\textit{For each instrument mentioned, we asked the following questions:}
\begin{itemize}[left=19pt]
        \item[\textbf{[IQ7]}] What type(s) of representational harms are you measuring with [this instrument]? 
        \item[\textbf{[IQ8]}] How did you decide to use [this instrument]? 
        \item[\textbf{[IQ9]}] How do you use [this instrument] in your evaluation(s)? 
        \item[\textbf{[IQ10]}] Where did [this instrument] come from? Did you develop it yourself, modify an existing [instrument], or use an existing [instrument] as-is? 
        \end{itemize}
\textit{If applicable, for one instrument that the interviewee developed themselves, we asked the following questions:}
    \begin{itemize}[left=19pt]
        \item[\textbf{[IQ11]}] Why did you decide to develop [this instrument] yourself?  
        \item[\textbf{[IQ12]}] What, if any, actions have you taken or plan to take upon seeing the measurements obtained using [this instrument]? 
    \end{itemize}
\textit{If applicable, for one instrument that the interviewee adapted from an existing instrument, we asked the following questions:}
    \begin{itemize}[left=19pt]
        \item[\textbf{[IQ13]}] Why did you decide to start with this existing [instrument]?  
        \item[\textbf{[IQ14]}] Why did you decide to modify [this instrument] rather than using it as-is? 
        \item[\textbf{[IQ15]}] What, if any, actions have you taken or plan to take upon seeing the measurements obtained using [this instrument]? 
    \end{itemize}
\textit{If applicable, for one instrument that the interviewee used as-is, we asked the following questions:}
    \begin{itemize}[left=19pt]
        \item[\textbf{[IQ16]}] Why did you decide to use this existing [instrument] as-is?  
        \item[\textbf{[IQ17]}] What, if any, actions have you taken or plan to take upon seeing the measurements obtained using [this instrument]? 
    \end{itemize}
\subsection{Challenges with measurement instruments for representational harms [15 min]}
    \begin{itemize}[left=19pt]
    \item[\textbf{[IQ18]}] Were there any other existing, publicly available [instruments] that you investigated using instead? 
        \item[\textbf{[IQ19]}] \textit{For each instrument mentioned:} Why did you decide not to use [this instrument]? 
    \item[\textbf{[IQ20]}] \textit{For each of the challenges defined below, say either:} 
    \item[] ``It sounds like you mentioned an issue to do with [challenge]. Is that correct?”, \textit{or} 
    \item[] “I don’t think you mentioned [challenge]. Did you experience any issues with this?'' 
    \item[] \textit{We provided interviewees with the following set of challenges related to measurement instruments:}
    \begin{itemize}[left=0pt]
    \item[--] Whether it is sufficiently specific to the system being evaluated and its particular use cases and deployment contexts
    \item[--] Whether it can be adapted for different systems, use cases, and deployment contexts
    \item[--] Whether it results in valid measurements – i.e., meaningfully measures what stakeholders think it measures 
    \item[--] Whether it results in similar measurements when used in similar ways, especially over time 
    \item[--] Whether its resulting measurements can be understood by stakeholders
    \item[--] Whether its resulting measurements can be acted upon by stakeholders
    \item[--] Whether it can scale to increasing workloads
    \end{itemize}
        \item[\textbf{[IQ21]}] \textit{For each challenge experienced:}  What, if anything, did you do to address this issue? 
    \item[\textbf{[IQ22]}] Did you experience any other issues that we haven’t discussed?  
        \item[\textbf{[IQ23]}] \textit{If applicable:} What, if anything, did you do to address this issue? 
    \end{itemize}

\subsection{Comparing measurement of representational harms to other harms [5 min]}
\begin{itemize}[left=19pt]
    \item[\textbf{[IQ24]}] Do your previous, current, or planned evaluation(s) of LLM-based system(s) involve measuring harms, adverse impacts, or other undesirable behaviors other than representational harms?  
    \item[\textbf{[IQ25]}] \textit{If yes:} What types of harms, adverse impacts, or other undesirable behaviors?  
    \item[\textbf{[IQ26]}] \textit{If yes:} Are your experiences measuring these types of harms, adverse impacts, or other undesirable behaviors similar to your experiences measuring representational harms? What, if anything, is similar and what, if anything, is different about your experiences? I'm especially interested in hearing about how the [instruments] you use to measure these types of harms, adverse impacts, or other undesirable behaviors are similar to or different from the [instruments] you use to measure representational harms. 
    \end{itemize}\subsection{Desired improvements to measuring representational harms [5 min]} 
\begin{itemize}[left=19pt]
    \item[\textbf{[IQ27]}] Putting aside any time or budget constraints, what, if anything, would you improve about the way that you previously, currently, or plan to measure representational harms?   
    \item[\textbf{[IQ28]}] What do you need, that you don't currently have, in order to make those improvements? 
\end{itemize}

\subsection{Closing [5 min]}
\begin{itemize}[left=19pt]
    \item[\textbf{[IQ29]}] Is there anything else you would like to tell us about your previous, current, or planned evaluation(s) of LLM-based system(s)?
\end{itemize}

\vspace{-0.25cm}
\bibliographystyle{plainnat}
\bibliography{references}

%%%%%%%%%%%%%%%%%%%%%%%%%%%%%%%%%%%%%%%%%%%%%%%%%%%%%%%%%%%%

%%%%%%%%%%%%%%%%%%%%%%%%%%%%%%%%%%%%%%%%%%%%%%%%%%%%%%%%%%%%

\end{document}